*Article*

# Extending the Technology Acceptance Model 3 to Incorporate the Phenomenon of Warm-Glow

**Antonios Saravanos \*, Stavros Zervoudakis and Dongnanzi Zheng**

Division of Applied Undergraduate Studies, New York University, New York, NY 10003, USA
\* Correspondence: saravanos@nyu.edu; Tel.: +1-212-992-8725

**Abstract:** In this paper, we extend the third evolution of the Technology Acceptance Model (TAM3) to incorporate warm-glow with the aim of understanding the role this phenomenon plays on user adoption decisions. Warm-glow is the feeling of satisfaction or pleasure (or both) that is experienced by individuals after they do something "good" for their fellow human. Two constructs—perceived extrinsic warm-glow (PEWG) and perceived intrinsic warm-glow (PIWG)—were incorporated into the TAM3 model to measure the two dimensions of user-experienced warm-glow, forming what we refer to as the TAM3 + WG model. An experimental approach was taken to evaluate the suitability of the proposed model (i.e., TAM3 + WG). A vignette was created to present users with a hypothetical technology designed to evoke warm-glow in participants. Our TAM3 + WG model was found to be superior in terms of fit to the TAM3 model. Furthermore, the PEWG and PIWG constructs were confirmed to be unique within the original TAM3 model. The findings indicate that the factors that have the greatest influence on consumer decisions are (in decreasing order) perceived usefulness (PU), PIWG, subjective norm (SN), and PEWG. Additionally, a higher PEWG resulted in the technology being perceived as more useful. In other words, both extrinsic and intrinsic warm-glow play a prominent role in user decisions as to whether or not to adopt a particular technology.

**Keywords:** warm-glow; technology adoption; TAM3

## 1. Introduction

Since the inception of the first technology acceptance model (TAM) developed by Davis [1], the exercise of modeling user adoption of technology has continued to evolve, emerging as a prime area of study in the field of information system management [2]. Given the prominent—and ever-increasing—role that technology now plays in the activities of individuals and organizations, this practice has become more widespread. Accordingly, organizations find value in the ability to ascertain whether a technology will be accepted by prospective users, as this can afford those organizations a competitive advantage [3]. Technology adoption modeling can trace its origins to the work of Fishbein and Ajzen [4] and their theory of reasoned action, and later to the theory of planned behavior [5,6], both of which were designed to predict the behavioral intentions of consumers with respect to adoption. However, these were developed to explain the behavior of people in general and not in a particular context. Over time, specialized models were developed specifically for use with technology, and two main strands emerged: the TAM and the unified theory of acceptance and use of technology (UTAUT) line of models [7]. The TAM line emerged from Davis's [1] attempt to adopt the Theory of Reasoned Action specifically for use with technology products. It is currently in its third evolution [2] and is referred to colloquially as TAM3. The UTAUT model was the result of a comprehensive review and subsequent synthesis of several models that emerged over time as alternatives to TAM. For this work, we will focus on the TAM3 model for several reasons: TAM is described as "an established approach in research on the acceptance of new technologies" [8], is "somewhat of a gold standard" [9], and is more widely used than UTAUT [10].

The TAM model (TAM0) emerged in 1986 as a model and corresponding instrument designed "to predict the likelihood of a new technology being adopted within a group or an organization" [11]. Accordingly, employers could use it to determine whether a new technology was going to be accepted and used by their employees. It was based on the idea that the perceived use and ease of use of a technology impact a user's attitude regarding that technology and subsequently their behavioral intention to use that technology. The first revision (TAM1) in 1989 saw the elimination of the attitude factor [12]. The second revision (TAM2) [13] in 2000 incorporated supplementary factors reflecting "social influence processes" as well as "cognitive instrumental processes" [14]. The third revision (TAM3) in 2008 found them extending the model even further by developing the antecedents with respect to the perceived effort required to use a technology [2]. While originally for use in understanding the adoption of technology in the workplace, the TAM3 instrument saw the perspective change to that of the individual. Importantly, these models and their respective instruments were designed to be flexible. TAM3 has been extended and adapted for a wide variety of cases to meet the needs of different technologies and the contexts in which they are used [15,16]. However, one aspect that is only now being explored within the context of the TAM3 model is the phenomenon of warm-glow.

One can trace the origins of the term "warm-glow" to the work of Andreoni [17], who reports that consumers may perceive such a feeling subsequently to having donated to the less fortunate and, in consequence, having "done their bit" for humanity. There are two dimensions to warm-glow which depend on the motivation of the consumer action: extrinsic and intrinsic. The first form, extrinsic warm-glow (EWG), represents the feeling derived by consumers for engaging in selfish (non-altruistic) behavior [18]. Andreoni [17] explains that "people have a taste for giving: perhaps they receive status or acclaim". The second, intrinsic warm-glow (IWG), represents the feeling derived by consumers when engaging in altruistic behavior [19], Saito described it as "a willingness to benefit others, even at one's own expense" [20]. We can see this feeling being evoked in consumers when making decisions for a form of technology which Saravanos et al. [21] categorize as "good tech" (i.e., technology products that are perceived by users as "good", and accordingly, this perception of goodness evokes in them a feeling of warm-glow). It should be noted that "good tech" is a perceptive category, which "is a subjective, adjustable category" [22]. Therefore, what may appear to one person as "good" may not appear that way for another. For example, a consumer who has a passion for the environment would regard the Ecosia product (a web-based search tool whose proceeds are primarily used for the planting of trees) to be "good tech" and would accordingly experience a feeling of warm-glow [21]. To capture user perception of the aforementioned dimensions of warm-glow, we can consider the work of Saravanos et al. [21], who offer constructs specifically developed with technology adoption modeling in mind. The first is "perceived extrinsic warm-glow" (PEWG), designed to measure user perception of EWG; the second is "perceived intrinsic warm-glow" (PIWG), designed to measure user perception of IWG [21]. These constructs include an accompanying instrument.

The purpose of this study is to extend (and evaluate) the TAM3 model for the warm-glow phenomenon, thereby offering insight as to the effect it (i.e., warm-glow) plays on end-user adoption decisions for "good tech". Accordingly, in this paper, we perform the following: (1) incorporate the warm-glow constructs (i.e., PEWG and PIWG) to measure the two dimensions (i.e., extrinsic and intrinsic) of warm-glow into the TAM3 model; (2) validate this new enhanced model which we shall refer to as TAM3 + WG; (3) confirm that the warm-glow constructs do not replicate the role of other (potentially duplicative) factors within the original TAM3 model; and (4) ascertain the relative magnitude of the effect that warm-glow plays in users' decisions to adopt technology they perceive to be "good", in comparison to the original TAM3 factors.

We find our proposed TAM3 + WG to be superior to the TAM3 model in terms of fit, with the PEWG and PIWG constructs being unique (i.e., not duplicating any of the original constructs within the model). Moreover, after consumer perception of a technology's usefulness, the perception of IWG represents the second-largest effect in consumer decisions. This is followed by the subjective norm held by a user and then by their perception of EWG. Thus, we establish that within our cultural context (i.e., the United States), IWG plays a greater role in consumer decision than EWG does. Furthermore, the perception of EWG leads to the user viewing the technology as more useful. Therefore, the research highlights the positive effect that warm-glow can have on consumers, which further justifies the inclusion of an element of "goodness" in technology products. Additionally, our work provides a model that can be used by those that seek to understand or predict the adoption of technology. Given that technology acceptance models are confirmatory by nature, the TAM3 + WG is to be used in cases where there is a realistic belief that a technology will evoke a feeling of warm-glow in a certain population.

## 2. Materials and Methods

In this section we describe the development of the hypothesis and model, followed by the data collection process that was utilized.

### 2.1. Hypothesis and Model Development

Given that technology acceptance modeling relies on a confirmatory approach, our first step is to establish the model that we are proposing for consideration (illustrated in Figure 1). We begin with the inclusion of the relevant constructs from Venkatesh and Bala's [2] original TAM3 model. At the core of the TAM3 model are three fundamental constructs that serve as determinants of customer acceptance of technology, referred to as "behavioral intention" (BI). These are "perceived ease of use" (PEOU), "perceived usefulness" (PU), and "subjective norm" (SN) [2]. BI is defined as "the degree to which a person has formulated conscious plans to perform or not perform some specified future behavior" [23]. The first of the determinants, PEOU, measures the effort that consumers perceive they will need to expend in order to use a particular technology, which we formally accept as "the degree to which a person believed that using a particular system would be free of effort" [12]. The second element, PU, measures the value that consumers perceive they will gain from using technology, which we regard as "the degree to which a person believes that using a particular system would enhance his or her job performance" [12]. The third, according to Venkatesh and Bala [2], is SN, which is "the degree to which an individual perceives that most people who are important to him think he should or should not use the system". To these, we add two supplemental constructs proposed by Saravanos et al. [21] to reflect the two dimensions of end-user-perceived warm-glow.

#### 2.1.1. Extrinsic Warm-Glow

It has been shown that warm-glow influences the adoption of what we define as "good technology", with the literature focusing primarily on environmentally sustainable technologies. With respect to the EWG aspect, the effects are illustrated in the work of Griskevicius, Tybur, and Van den Bergh [24], who observe that the use of green products is a way for some consumers to signal to others that they are affluent enough to consume products that have a positive benefit for society (and the environment), even though the products themselves may be of lower quality. Another similar study is offered by Griskevicius and Tybur [25], who illustrate that consumers' quest for status can lead to the purchase of products that are priced higher than their non-green counterparts. They draw the conclusion that consumer selection is not always made on merits such as quality, environmental benefits, or price, but rather because in contemporary western society, doing good often has a higher effect on an individual's image than luxury does. Similarly in Dastrup et al.'s [26] article, the authors investigate home electricity generation through solar panels and its effect on social status. While the authors do not explicitly link to the concept of (extrinsic) warm-glow and its perception, nor do they look explicitly at individual user perceptions, they do observe an effect. These examples justify incorporating a construct into our model to reflect consumer perceptions of EWG, "perceived extrinsic warm-glow" (PEWG), and the proposal of the following hypothesis:

**H1:** *Perceived extrinsic warm-glow (PEWG) positively influences behavioral intention (BI).*

2.1.2. Intrinsic Warm-Glow

With respect to the IWG aspect, there are several studies which illustrate its effect on user adoption. Hartmann and Apaolaza-Ibáñez [27] examine the attitudes and intentions of consumers with regard to green energy brands. One of the aspects they examine is that of (intrinsic) warm-glow incorporating a corresponding construct, which reveals that it does increase consumer purchasing behavior. Ma and Burton [28] explore consumer decisions to purchase green electricity in Australia. The authors find that warm-glow does influence consumer decisions more than the actual attributes of the competing products. More recently, we see these ideas applied in the work of Sun et al. [29] who utilize the "Theory of Planned Behavior" [5,6] to explore the attitude displayed by consumers with regard to their intention to purchase the installation of rooftop solar photovoltaic (PV) systems in Taiwan. Their warm-glow construct focuses exclusively on the intrinsic dimension of warm-glow, once again demonstrating how it can influence consumer attitudes, this time towards the installation of rooftop PV's. Azalia et al. [30] build on this work, investigating how individual concerns for the environment, warm glow, and financial factors influence the adoption of solar PV's in Indonesia. In their paper, they like Sun et al. [29], also rely on the "Theory of Planned Behavior" [5,6] and focus on the intrinsic dimension. They likewise find (intrinsic) warm-glow to have a statistically significant effect "in the motivation of using solar PV" [30]. Another example is offered in the work of Bhutto et al. [31], who look at the adoption of energy-efficient home appliances (EEAs) in Pakistan (again by extending the Theory of Planned Behavior). The authors conclude that "warm glow benefits motivate consumers to pay premium prices for EEAs to feel moral satisfaction". Tangentially, we can also look to the work of Karjalainen and Ahvenniemi [32]. They apply Tiger's [33] framework to investigate the pleasure (specifically categorized as physical, social, psychological, and ideological) derived by those who adopt PV's in Finland. Karjalainen and Ahvenniemi [32] describe (ideological) pleasure being derived from "the capability to produce [one's] own clean energy and reduce emissions", as well as "the ability to provide clean energy for other energy users". In essence, they are referring to the intrinsic dimension of the warm-glow phenomenon. Accordingly, we incorporate a factor to reflect IWG perceptions held by consumers into our model, "perceived intrinsic warm-glow" (PIWG), and subsequently consider the following hypothesis:

**H2:** *Perceived intrinsic warm-glow (PIWG) positively influences behavioral intention (BI).*

2.1.3. The Influence of Warm-Glow on PEOU and PU

The existing literature is not explicit about the kind of influence the warm-glow phenomenon can have on the main antecedents of consumer BI (specifically, PEOU and PU). However, studies have shown that external factors can influence the aforementioned antecedents of BI. These factors include a consumer's image, anxiety about using technology, price, privacy, and trust [34,35]. Therefore, for completeness, we further postulate that such a relationship could exist between the two primary determinants of BI; PU; PEOU; and the warm-glow constructs, PEWG and PIWG. Consequently, we propose the following hypotheses:

**H3:** *Perceived extrinsic warm-glow (PEWG) positively influences perceived ease of use (PEOU).*

**H4:** *Perceived intrinsic warm-glow (PIWG) positively influences perceived ease of use (PEOU).*

**H5:** *Perceived extrinsic warm-glow (PEWG) positively influences perceived usefulness (PU).*

**H6:** *Perceived intrinsic warm-glow (PIWG) positively influences perceived usefulness (PU).*

2.1.4. Determining the Uniqueness of the PEWG and PIWG Constructs

Finally, we must account for the possibility that there may be factors in the original TAM3 model that could act as substitutes to the constructs of PEWG and PIWG. With respect to EWG, there are two factors in the current TAM model that could potentially serve as substitutes to our PEWG construct. The first is "image" (IMG), which Moore and Benbasat [35] define as "the degree to which an individual perceives that use of an innovation will enhance his or her status in his or her social system". The second is SN. Therefore, we propose the following hypotheses:

**H7:** *Perceived extrinsic warm-glow (PEWG) serves as a substitute to image (IMG) with respect to perceived usefulness (PU).*

**H8:** *Perceived extrinsic warm-glow (PEWG) serves as a substitute to subjective norm (SN) with respect to perceived usefulness (PU).*

For IWG, there are two constructs in the model that measure pleasure and serve as possible competitors to PIWG. The first is "perceived enjoyment" (ENJ), which Venkatesh [36] defines as "the extent to which the activity of using a specific system is perceived to be enjoyable in its own right, aside from any performance consequences resulting from system use". Although a connection between the two constructs is not well established, we can see hints of a relationship in the work of Kuruvatti et al. [37], who report "fun" to be one of the reasons people donate blood. The second factor to consider is that of "computer playfulness" (CPLAY), which Webster and Martocchio [38] define as "the degree of cognitive spontaneity in microcomputer interactions". To investigate the relationship between perceived IWG and the existing factors of ENJ and CPLAY, we establish the following hypotheses:

**H9:** *Perceived intrinsic warm-glow (PIWG) is a substitute to perceived enjoyment (ENJ) with respect to perceived ease of use (PEOU).*

**H10:** *Perceived intrinsic warm-glow (PIWG) is a substitute to computer playfulness (CPLAY) with respect to perceived ease of use (PEOU).*

In addition to the variable PU, the factor of SN is also associated in the TAM through the variable BI. For completeness, it follows that we additionally propose this final hypothesis as well:

**H11:** *Perceived extrinsic warm-glow (PEWG) is a substitute to subjective norm (SN) with respect to behavioral intention (BI).*

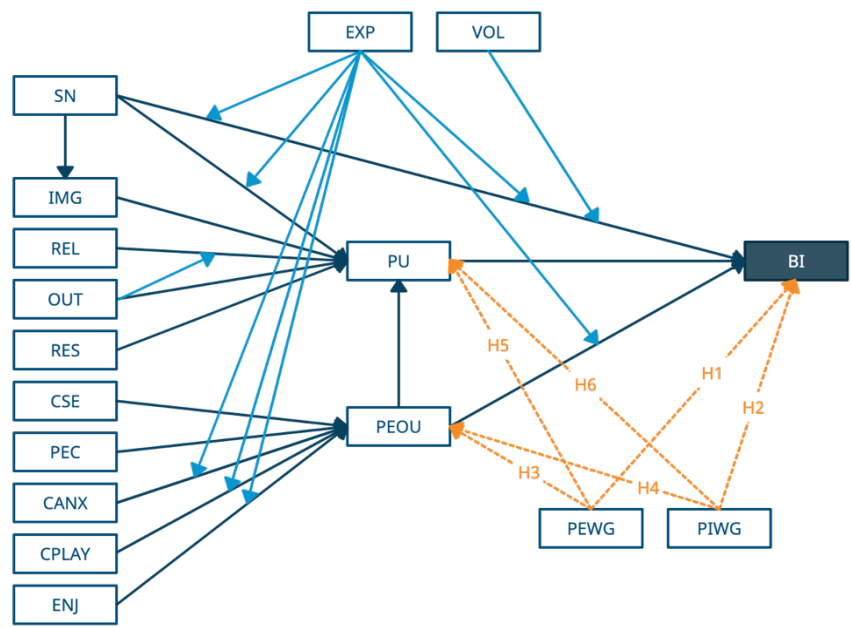

**Figure 1.** Illustration of our proposed model.

*2.2. Data Collection*

To evaluate the proposed model participants were presented with a hypothetical internet search engine that simulated the presence of extrinsic and intrinsic warm-glow, as described by Saravanos et al. [21]. All participants were first asked to confirm their willingness to participate. Next, those that chose to continue answered questions with respect to their gender, age, income, schooling, race, and prior experience with internet search technology. Subsequently, they were presented with a vignette that described a hypothetical technology product (see Saravanos et al. [21]). They then filled out a questionnaire designed to assess user perception of the product. This included the respective questions from Venkatesh and Bala's [2] TAM3 instrument, which were adjusted for consumers rather than for a workplace context, and specifically for the case of internet search technology, as well as the warm-glow questions from Saravanos et al. [21]. Lastly, we incorporated two questions from Abbey and Meloy [39], which we modified to gauge participant attention. All questions—those from the TAM3 instrument, the PIWG and PEWG constructs, and the attention check questions—were rated using a 7-point Likert scale ranging from "strongly disagree" to "strongly agree". The vignettes and questionnaire were distributed through the use of the Qualtrics' online survey platform.

We recruited our participants using the Amazon Mechanical Turk crowdsourcing platform, which has been used previously for such studies [40]. To identify an appropriate minimum size for our sample we looked to the "10 times rule method", which is "the most widely used minimum sample size estimation method in PLS-SEM" [41]. This approach states that the sample size "should be equal to", "10 times the largest number of formative indicators used to measure a single construct" [42]. Given we have, in the most conservative case, 8 formative indicators (i.e., for the BI construct), the sample should be over 80. We also recognize that while SEM-PLS is well suited for small sample sizes, researchers recognize the advantage of larger sample sizes. For example, Chin and Newsted [43] recommend samples be above 150, which they then go on to describe as large. Accordingly, we collected a total of 405 responses, of which all participants were from the United States. Of those, 80 submissions were removed from the final dataset, as they failed to pass the attention checks or were incomplete, leaving 325 remaining responses. The breakdown can be seen in greater detail in Table 1. Almost all participants appeared to be frequent users of search technology, with 96.92% indicating that they use a search engine on a daily basis. The majority of participants (83.38%) indicated that their favorite search engine was Google, followed by DuckDuckGo (10.77%) and Bing (3.69%).

**Table 1.** Demographic Profile of Respondents.

| Item | Type | Frequency (*n* = 325) | Percentage (%) |
|---|---|---|---|
| Gender | Female | 128 | 39.38 |
| | Male | 193 | 59.38 |
| | Other | 1 | 0.31 |
| | Prefer not to answer | 3 | 0.92 |
| Age | 18–25 | 16 | 4.92 |
| | 26–30 | 54 | 16.61 |
| | 31–35 | 80 | 24.62 |
| | 36–45 | 91 | 28.00 |
| | 46–55 | 51 | 15.69 |
| | 56 or older | 30 | 9.23 |
| | Prefer not to answer | 3 | 0.92 |
| Income | Less than $10,000 | 9 | 2.77 |
| | $10,000 to $19,999 | 27 | 8.31 |
| | $20,000 to $29,999 | 29 | 8.92 |
| | $30,000 to $39,999 | 45 | 13.85 |
| | $40,000 to $49,999 | 32 | 9.85 |
| | $50,000 to $59,999 | 41 | 12.62 |
| | $60,000 to $69,999 | 35 | 10.77 |
| | $70,000 to $79,999 | 29 | 8.92 |
| | $80,000 to $89,999 | 12 | 3.69 |
| | $90,000 to $99,999 | 12 | 3.69 |
| | $100,000 to $149,999 | 28 | 8.62 |
| | $150,000 or more | 19 | 5.85 |
| | Prefer not to answer | 7 | 2.15 |
| Schooling | Less than high school degree | 4 | 1.23 |
| | High school graduate (high school diploma or equivalent including GED) | 41 | 12.62 |
| | Some college but no degree | 67 | 20.62 |
| | Associate degree in college (2-year) | 32 | 9.85 |
| | Bachelor's degree in college (4-year) | 136 | 41.85 |
| | Master's degree (e.g., MA, MS) | 31 | 9.54 |
| | Professional degree (e.g., MBA, MFA, JD, MD) | 6 | 1.85 |
| | Doctoral degree (e.g., PhD, EdD, DBA) | 5 | 1.54 |
| | Prefer not to answer | 3 | 0.92 |

## 3. Results

In this section, we outline the analysis undertaken and then report on the results.

### 3.1. Measurement Model

In the first stage, a measurement model was used to examine the relationship between the manifest variables and their corresponding latent variables. This was done to ascertain whether the manifest variables effectively measured the latent variables. To accomplish this, we assessed the measures of convergent validity, construct reliability, and discriminant validity using SmartPLS3.3.2, SmartPLS GmbH, Germany [44]. To test convergent validity, which reveals how items are positioned between reality and theory [45], we relied on the factor loadings and the average variance extracted (AVE). We removed any manifest variables with values lower than 0.7 from our model for both of these criteria, as prescribed by Chin [46]. Specifically, we removed in sequence, VOL1 (0.290), RES4 (0.440), PEC4 (0.488), CPLAY4 (0.569), and PEC1 (0.663). The remaining items were statistically significant ($p < 0.05$, $t$-statistics were obtained from bootstrapping with 7000 subsamples), reflecting that they possessed appropriate convergent validity (see Table 2). To test construct reliability, we used the measures of composite reliability (CR) and Cronbach's Alpha. For both measures, we found values greater than 0.7 indicating overall good construct reliability, except for VOL, which had a Chronbach's Alpha of 0.548 and a CR score of 0.556, representative of acceptable construct reliability (see Table 3). Finally, to test discriminant validity, we used the Fornell–Larcker criterion as well as cross-loadings. With regard to the Fornell–Larcker criterion, Fornell and Larcker [47] advise that the correlations between each construct should be lower than the square root of the AVE. Concerning the cross-loadings, Chin [48] advises that each cross-loading be lower than all of the indicator's loadings. Since we satisfied both requirements, we concluded that our measurement model's discriminant validity was satisfactory. Given that our TAM3 + WG model had acceptable convergent validity as well as suitable reliability and discriminant validity we felt confident to apply the manifest variables in order to investigate the concurrent validity and sensitivity of our warm-glow constructs, as well as the ensuing structural model.

**Table 2.** Summary of Convergent Validity Testing.

| Factor | Item | Loading | t-Statistic | AVE |
|---|---|---|---|---|
| BI | BI2 | 0.970 | 232.177 * | 0.938 |
| | BI3 | 0.967 | 191.694 * | |
| CANX | CANX1 | 0.920 | 53.383 * | 0.850 |
| | CANX4 | 0.924 | 66.114 * | |
| CPLAY | CPLAY1 | 0.830 | 6.984 * | 0.796 |
| | CPLAY2 | 0.933 | 6.029 * | |
| | CPLAY3 | 0.911 | 6.500 * | |
| CSE | CSE1 | 0.792 | 23.705 * | 0.670 |
| | CSE2 | 0.858 | 46.671 * | |
| | CSE3 | 0.789 | 24.723 * | |
| | CSE4 | 0.834 | 29.711 * | |
| ENJ | ENJ2 | 0.974 | 270.531 * | 0.942 |
| | ENJ3 | 0.967 | 145.771 * | |
| IMG | IMG1 | 0.923 | 55.107 * | 0.881 |
| | IMG2 | 0.944 | 92.759 * | |
| | IMG3 | 0.948 | 129.774 * | |
| OUT | OUT2 | 0.962 | 120.328 * | 0.929 |
| | OUT3 | 0.966 | 169.197 * | |
| PEC | PEC2 | 0.905 | 48.966 * | 0.834 |
| | PEC3 | 0.922 | 41.606 * | |
| PEOU | PEOU1 | 0.928 | 55.903 * | 0.792 |
| | PEOU2 | 0.749 | 15.346 * | |
| | PEOU3 | 0.937 | 84.092 * | |
| | PEOU4 | 0.931 | 81.875 * | |
| PEWG | PEWG1 | 0.916 | 77.634 * | 0.838 |
| | PEWG2 | 0.910 | 81.200 * | |
| | PEWG3 | 0.920 | 73.172 * | |
| PIWG | PIWG1 | 0.941 | 76.172 * | 0.883 |
| | PIWG2 | 0.930 | 71.715 * | |
| | PIWG3 | 0.948 | 115.057 * | |
| PU | PU3 | 0.910 | 63.337 * | 0.836 |
| | PU4 | 0.919 | 82.399 * | |
| | REL1 | 0.936 | 36.904 * | |
| REL | REL2 | 0.950 | 76.491 * | 0.890 |
| | REL3 | 0.944 | 86.483 * | |
| | RES1 | 0.871 | 30.994 * | |
| RES | RES2 | 0.803 | 18.491 * | 0.731 |
| | RES3 | 0.887 | 44.698 * | |
| | SN1 | 0.927 | 92.262 * | |
| SN | SN2 | 0.925 | 65.683 * | 0.792 |
| | SN3 | 0.841 | 37.458 * | |
| | SN4 | 0.864 | 49.555 * | |
| VOL | VOL2 | 0.868 | 7.697 * | 0.687 |
| | VOL3 | 0.787 | 5.128 * | |

* $p < 0.01$.

**Table 3.** Summary of Reliability Testing.

| Factor | Number of Items | Cronbach's Alpha | CR |
|---|---|---|---|
| BI | 2 | 0.934 | 0.968 |
| CANX | 2 | 0.823 | 0.919 |
| CPLAY | 3 | 0.879 | 0.921 |
| CSE | 4 | 0.836 | 0.890 |
| ENJ | 2 | 0.939 | 0.970 |
| IMG | 3 | 0.932 | 0.957 |
| OUT | 2 | 0.924 | 0.963 |
| PEC | 2 | 0.801 | 0.909 |
| PEOU | 4 | 0.911 | 0.938 |
| PEWG | 3 | 0.904 | 0.940 |
| PIWG | 3 | 0.934 | 0.958 |
| PU | 2 | 0.804 | 0.911 |
| REL | 3 | 0.938 | 0.960 |
| RES | 3 | 0.815 | 0.890 |
| SN | 4 | 0.912 | 0.938 |
| VOL | 2 | 0.548 | 0.814 |

*3.2. Structural Model*

In the second stage, we employed partial least squares (PLS), which is a type of structural equation modelling (SEM), specifically to test our conceptual model (depicted in Figure 1), once again using SmartPLS3.3.2 [44]. The use of PLS-SEM has, according to Hair, Ringle, and Sarstedt [44], "been increasingly applied in marketing and other business disciplines", with the authors describing it as "a 'silver bullet' or panacea for dealing with empirical research challenges". PLS-SEM is an alternative to CB-SEM which Hair et al. [49] describe as the better-known approach, writing that "for many researchers, SEM is equivalent to carrying out covariance-based SEM (CB-SEM)". Rigdon et al. [50] go further and point out that "two opposing camps" exist and that there is controversy regarding the suitability of methodologies. Jannoo et al. [51] write that "CB-SEM requires a set of stringent assumptions, such as normality of data and adequate sample size". The authors go on to note that in cases where the CB-SEM assumptions are not satisfied PLS-SEM should be utilized, referencing the work of Haenlein and Kaplan [52] and Rigdon et al.

[53]. We elected to use the PLS-SEM approach over CB-SEM because it is better suited for our case and, in particular, able to "handle small sample sizes", "complex models with numerous endogenous and exogenous constructs and indicator variables", and "non-normal data distributions", as prescribed by Astrachan et al. [54].

First, we inspected the variance inflation factor (VIF) to appraise the level of collinearity of our latent variables. Hair et al. [55] writes, "VIF values of 5 or above indicate critical collinearity issues among the indicators of formatively measured constructs". Accordingly, any values greater than 5 should be removed (as per Ringle and Sarstedt [49]). Specifically, we removed, BI1 (16.074), ENJ1 (13.121), PU1 (11.549), OUT1 (7.598), CANX3 (10.353), CANX2 (6.634), and PU2 (6.281). Thereby, any concerns vis-à-vis collinearity with our data were alleviated [55,56]. We found that our model explained 67.7% of the BI of an individual, in terms of accepting the technology of web-based searches. The significant antecedents of the BI factor were (in order of decreasing strength) as follows: PU ($\beta$ = 0.296; $p < 0.01$), PIWG ($\beta$ = 0.220; $p < 0.01$), SN ($\beta$ = 0.211; $p < 0.01$), PEWG ($\beta$ = 0.190; $p < 0.01$), and VOL ($\beta$ = −0.106; $p < 0.01$). Therefore, each increase of 1 unit in PEWG led to an increase of 0.190 units in BI. Hence, the result was consistent with H1. Similarly, each increase of 1 unit in PIWG led to an increase of 0.220 units in BI, so the result was consistent with H2.

Furthermore, with respect to PEOU, the model explained 53.0% of the variance, with the significant factors (in order of decreasing strength) as follows: "perception of external control" (PEC) ($\beta$ = 0.359; $p < 0.01$), "computer anxiety" (CANX) ($\beta$ = 0.240; $p < 0.01$), ENJ ($\beta$ = 0.214; $p < 0.01$), and CSE ($\beta$ = 0.193; $p < 0.01$). Interestingly, both H3 and H4 were not verified. Therefore, we were unable to assume a relationship between the constructs of PEWG and PIWG with PEOU.

With respect to PU, the model explained 59.8%, with significant factors (in order of decreasing strength) of "output quality" (OUT) ($\beta$ = 0.252; $p < 0.01$), PEWG ($\beta$ = 0.195; $p < 0.01$), SN ($\beta$ = 0.150; $p < 0.05$), PEOU ($\beta$ = 0.139; $p < 0.01$), IMG ($\beta$ = 0.115; $p < 0.05$), and REL ($\beta$ = 0. 112; $p < 0.05$). Interestingly, H6 was not verified. Therefore, we were unable to assume a relationship between the constructs of PIWG and PU. With respect to H5, we did find a statistically significant relationship between PU and PEWG. Therefore, the results indicated that a higher perception of EWG led to a higher perception of usefulness by using the technology. Lastly, with respect to IMG, SN ($\beta$ = 0.544; $p < 0.01$) explained 29.5% of the variance. Table 4 summarizes the results from the structural model and Table 5 the results from the testing of the hypotheses. Table 6 lists all $R^2$ values.

**Table 4.** Structural Model Results.

| Path | $\beta$ | t-Statistic |
|---|---|---|
| EXP → BI | 0.101 | 1.836 |
| PEOU → BI | 0.023 | 0.603 |
| PEWG → BI | 0.190 | 2.906 ** |
| PIWG → BI | 0.220 | 3.424 ** |
| PU → BI | 0.296 | 5.221 ** |
| SN → BI | 0.211 | 3.345 ** |
| VOL → BI | −0.106 | 2.880 ** |
| EXP x PEOU → BI | 0.071 | 1.510 |
| EXP x SN → BI | −0.075 | 1.281 |
| PEWG x SN → BI | 0.015 | 0.529 |
| VOL x SN → BI | 0.052 | 1.546 |
| SN → IMG | 0.544 | 13.405 ** |
| CANX → PEOU | 0.240 | 4.061 ** |
| CPLAY → PEOU | −0.013 | 0.286 |
| CSE → PEOU | 0.193 | 2.652 ** |
| ENJ → PEOU | 0.214 | 3.456 ** |
| EXP → PEOU | −0.049 | 1.293 |
| PEC → PEOU | 0.359 | 4.624 ** |
| PEWG → PEOU | −0.070 | 1.143 |
| PIWG → PEOU | −0.049 | 0.775 |
| EXP x CANX → PEOU | 0.070 | 0.934 |
| EXP x CPLAY → PEOU | 0.003 | 0.057 |
| EXP x ENJ → PEOU | 0.006 | 0.076 |
| PIWG x CPLAY → PEOU | −0.070 | 1.091 |
| PIWG x ENJ → PEOU | 0.043 | 0.783 |
| EXP → PU | 0.052 | 1.126 |
| IMG → PU | 0.115 | 2.154 * |
| OUT → PU | 0.252 | 4.304 ** |
| PEOU → PU | 0.139 | 2.900 ** |
| PEWG → PU | 0.195 | 3.036 ** |
| PIWG → PU | 0.025 | 0.333 |
| REL → PU | 0.112 | 2.012 * |
| RES → PU | 0.043 | 0.873 |
| SN → PU | 0.150 | 2.195 * |
| EXP x PEOU → PU | 0.036 | 0.963 |
| EXP x SN → PU | −0.079 | 1.520 |
| OUT x REL → PU | 0.011 | 0.377 |
| PEWG x IMG → PU | −0.043 | 0.979 |
| PEWG x SN → PU | −0.082 | 1.872 |

* $p < 0.05$; ** $p < 0.01$.

**Table 5.** Hypothesis Testing Results.

| Hypothesis | Relationship | Decision |
|---|---|---|
| H1 | PEWG → BI | Supported |
| H2 | PIWG → BI | Supported |
| H3 | PEWG → PEOU | Not Supported |
| H4 | PIWG → PEOU | Not Supported |
| H5 | PEWG → PU | Supported |
| H6 | PIWG → PU | Not Supported |
| H7 | PEWG x IMG → PU | Not Supported |
| H8 | PEWG x SN → BI | Not Supported |
| H9 | PIWG x ENJ → PEOU | Not Supported |
| H10 | PIWG x CPLAY → PEOU | Not Supported |
| H11 | PEWG x SN → PU | Not Supported |

**Table 6.** $R^2$ Values.

| Factor | $R^2$ |
|---|---|
| BI | 0.677 |
| IMG | 0.295 |
| PEOU | 0.530 |
| PU | 0.598 |

The existence of a substitutive relationship was considered by using moderators, following the approach of Hagedoorn and Wang [57] (see Tables 4 and 5). We found no statistically significant moderating role between PEWG and SN with respect to the dependent variable BI or between PIWG and ENJ or PIWG and CPLAY for the dependent variable

PEOU. For the dependent variable PU, we found no statistically different moderating role between PEWG and SN or PEWG and IMG. Therefore, we were able to discount the possibility that the aforementioned independent variables could serve as substitutes for one another, supporting H7 to H11. This allowed us to conclude that the PIWG and PEWG constructs were unique to the model and enabled it to capture a new phenomenon. In short, within the TAM3 + WG model, we found no substitutive relationship between PEWG and PIWG with existing similar constructs.

*3.3. Explanatory Power, Predictive Ability, and Model Fit for BI*

Concerning the dependent variable BI, the TAM3 + WG model had an $R^2$ of 0.677, which can be described as substantial (as defined by Hair, Ringle, and Sarstedt [49]). We also explored whether the addition of the PIWG or PEWG independent variables influenced $R^2$ within the model and examined whether those models were superior to our TAM3 + WG model, which was 5.1% higher than the original TAM3 model (see Table 7). The findings revealed that for the dependent variable BI, the associated independent variables in the TAM3 + WG model explained a larger proportion of the variance (when compared with the original TAM3 model, the TAM3 model with the PIWG construct, and the TAM3 model with PEWG construct). Consequently, we were able to conclude that the TAM3 + WG model had superior explanatory power. With respect to the predictive ability of the TAM3 + WG model, as the $Q^2$ for BI was greater than 0 (see Table 7), we can assert that the latent factors associated with BI do indeed have predictive ability [58]. With regard to model fit, all of the models in Table 7 had a standardized root mean square residual (SRMR) value of less than 0.08, which is the minimum acceptable value according to Hu and Bentler [59]. This further indicates that the TAM3 + WG model is a good fit [60,61]. Furthermore, of the two models compared, the TAM3 + WG model had the lower Akaike Information Criterion (AIC) value with respect to the dependent variable BI (AIC = −344.489). According to Akaike's [62] guidelines, this allows us to conclude that the TAM3 + WG model has the best fit.

**Table 7.** Comparison of Models for BI Factor.

|  | **TAM3** | **TAM3 + WG** |
|---|---|---|
| $R^2$ | 0.626 | 0.677 |
| $\Delta R^2$ |  | 0.051 |
| $Q^2$ | 0.570 | 0.614 |
| SRMR | 0.048 | 0.047 |
| AIC | −302.727 | −344.489 |

**4. Discussion and Conclusions**

Through this work we developed a new model (i.e., TAM3 + WG), in principle an extension of the popular TAM3 model, in order to incorporate warm-glow. The proposed TAM3 + WG, as well as the final validated version of the model, can be seen in Figures 1 and 2, respectively. Our aim being to appreciate the effect that warm-glow plays on consumer adoption decisions for "good tech" as defined by Saravanos et al. [21]. We incorporated the PEWG and PIWG constructs, proposed by Saravanos et al. [21], into the TAM3 model to form an enhanced model, TAM3 + WG, which now explicitly takes the warm-glow phenomenon into consideration. Our TAM3 + WG model was found to be superior to the TAM3 model in terms of fit when determining users' BI to accept "good tech". Moreover, none of the potentially competing factors in the existing TAM3 model were found to be appropriate substitutes to the constructs of PEWG and PIWG. The finding that warm-glow plays a prominent role in consumer decisions with regards to accepting technology and can influence those (consumer) decisions is not novel in and of itself; rather, it serves to support what has been suggested by earlier studies. Certainly, our work agrees with the findings of others who have explored the effect of EWG on technology in the past, such as Griskevicius, Tybur, and Van den Bergh [24]; Griskevicius and Tybur [25]; and Dastrup et al. [26]. In addition, our work corroborates studies that conclude that IWG influences adoption, such as those presented by Hartmann and Apaolaza-Ibáñez [27], Ma and Burton [28], Karjalainen and Ahvenniemi [32], Sun et al. [29], Azalia et al. [30], and Bhutto et al. [31].

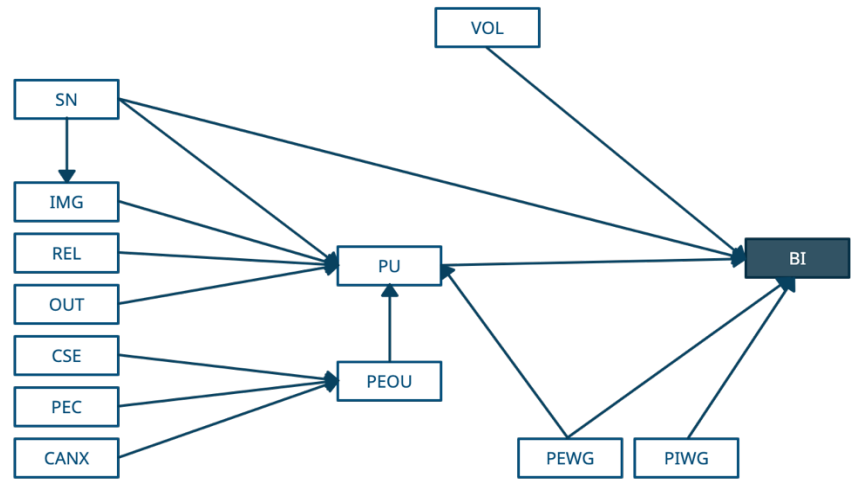

**Figure 2.** Illustration of our validated model.

Furthermore, our work offers insight into the relative magnitude of the warm-glow effect with respect to "traditional" constructs that affect the adoption of technology products. Correspondingly, PIWG represents the second-largest effect ($\beta = 0.220$) in consumer decisions, preceded only by their PU of the technology ($\beta = 0.296$). PEWG played the fourth-greatest role in determining consumer decisions related to technology adoption ($\beta = 0.190$); this is slightly less than SN ($\beta = 0.211$), which played the third-greatest role. Hence, we find that PIWG plays a greater role in consumer decision than PEWG does. Moreover, the combined magnitude of effects that each form of warm-glow had on determining consumer intention was similar and comparable to that of the technology's PU. This would explain why warm-glow appears to play such a critical role in technology adoption decisions.

Our study also sought to ascertain whether warm-glow can influence a consumer's perception of how easy it will be to use a technology (reflected through the PEOU construct) and how valuable that technology will be (reflected through the PU construct). For the most part, we found that these factors, which traditionally serve as the primary antecedents of consumer BI to accept a technology, were not influenced by the presence of either form of warm-glow, aside from the case of PU and PEWG. In other words, if a user perceived that they would receive EWG from a technology, they would consequently also perceive it to be more useful.

*4.1. Implications*

This study contributes to the general technology adoption literature by being, to the best of our knowledge, the first empirical study to extend the TAM3 model with warm-glow through the use of the PEWG and PIWG constructs to study the category of "good tech". From the practical perspective, the results highlight the magnitude of influence that the respective forms of warm-glow have on user adoption decisions (i.e., second-and fourth-greatest roles). Moreover, it shows how the perceived usability of a technology can be influenced through warm-glow (in our case, how extrinsic warm-glow can influence the usefulness that one perceives will be gleaned from its use). This finding can justify the creation of "good tech" by providing evidence of the advantage it has over "traditional" technology. Indeed, it clearly demonstrates that for organizations, not only is "doing the right thing" ethical—it can also be strategic. Moreover, the findings offer insight to marketers of "good tech" as they identify the key factors that influence consumer decisions and assess how important, they are in those decisions. This is further informed by the finding that consumers (in the United States) value the intrinsic (i.e., altruistic) dimension over the extrinsic (i.e., non-altruistic).

*4.2. Limitations and Future Research Directions*

We conclude by highlighting the key limitations of this study, which also conjointly offer insight into how this research can developed in the future. The first limitation concerns the model that we selected to extend, TAM3, and the recognition that outside of this particular model there is a wide spectrum of different models available for the study of technology adoption [63]. In our work, we selected the TAM3 model for two reasons: firstly, the TAM line of models serves as the original line of dedicated technology adoption models; and secondly, they represent one of the most widely used technology adoption models over the years [15,16,64]. That being said, other technology adoption models are indeed available [63], the most popular being UTAUT [63] and its latest version, UTAUT2 [65]. While this study serves as a strong starting point, we recommend that future work goes on to examine the inclusion and evaluation of our proposed warm-glow constructs into these other models (such as UTAUT).

The second limitation concerns the possible influences that culture may play on consumer BI to accept technology. Indeed, there is a plethora of examples [66–71] within technology adoption literature in which authors share findings that reveal the influence of culture on consumer BI. Similarly, as pointed out by Saravanos et al. [21], social innovation literature reports that the extent to which warm-glow impacts user decisions varies based on the culture of those consumers [72]. Consequently, given that our data were collected exclusively in the United States, future research may be interested in investigating our model's suitability for use with other cultures and go on to explore possible methods of integrating cultural factors into the model.

The third limitation concerns the characteristics of the technology we used to test our model for this study: a generic (unbranded) internet search solution, which is a standard, easy-to-use technology that is frequently offered without charge. Products with different characteristics regarding familiarity, complexity, price, and brand may lead the model to behave differently. For example, with respect to price, if one were to consider a product that is not free—perhaps even something expensive, such as a hybrid automobile—this might result in a different interplay between the factors in our model. We also offer a second example with respect to brand. Suppose the popular free-search engine Google were to offer a separate search product to compete with Ecosia and use any profits generated exclusively by that new search product for the environment (e.g., "Google Green"). Ecosia is a free search engine that dedicates its profits for the environment. This new hypothetical Google product, given that it would be part of the Google brand, may lead consumers to act differently. Consequently, we would argue that research investigating the interplay between these factors and the presence of warm-glow would provide valuable insight into end-user technology adoption behavior.

The fourth limitation concerns the fact that our work looks exclusively at user behavioral intention to adopt and not whether people went on to exhibit actual usage behaviors. Accordingly, further work is required to give insight into how these behavioral intentions to adopt convert into such behaviors.

**Author Contributions:** Conceptualization, A.S.; data curation, A.S.; writing—original draft preparation, A.S., D.Z., and S.Z.; writing—review and editing, A.S., D.Z., and S.Z.; supervision, A.S.; project administration, A.S.; funding acquisition, A.S. All authors have read and agreed to the published version of the manuscript.

**Funding:** This research was funded in part by a New York University School of Professional Studies Dean's Research Grant.

**Institutional Review Board Statement:** The study was conducted in accordance with the Declaration of Helsinki and approved by the Institutional Review Board of New York University (protocol code IRB-FY2022-6281 approved on 3 February 2022).

**Informed Consent Statement:** Informed consent was obtained from all subjects involved in the study.

**Data Availability Statement:** The data that support the findings of this study are available from the corresponding author, A.S., upon reasonable request.

**Conflicts of Interest:** The authors declare no conflict of interest.

**References**

1. Davis, F. A Technology Acceptance Model for Empirically Testing New End-User Information Systems: Theory and Results. Ph.D. Thesis, Massachusetts Institute of Technology, Cambridge, MA, USA, 1985.
2. Venkatesh, V.; Bala, H. Technology Acceptance Model 3 and a Research Agenda on Interventions. *Decis. Sci.* **2008**, *39*, 273–315. https://doi.org/10.1111/j.1540-5915.2008.00192.x.
3. Al-Harby, F.; Qahwaji, R.; Kamala, M. The Effects of Gender Differences in the Acceptance of Biometrics Authentication Systems within Online Transaction. In Proceedings of the International Conference on CyberWorlds, Bradford, UK, 7 September 2009; pp. 203–210. https://doi.org/10.1109/cw.2009.40.
4. Fishbein, M.; Ajzen, I. *Belief, Attitude, Intention, and Behavior: An Introduction to Theory and Research*; Addison-Wesley: Reading, MA, USA, 1975.
5. Ajzen, I. From Intentions to Actions: A Theory of Planned Behavior. In *Action Control: From Cognition to Behavior*; Kuhl, J., Beckmann, J., Eds.; Springer: Berlin/Heidelberg, Germany, 1985; pp. 11–39, ISBN 978-3-642-69746-3.
6. Ajzen, I.; Fishbein, M. *Understanding Attitudes and Predicting Social Behavior*; Pearson: Englewood Cliffs, NJ, USA, 1980; ISBN 978-0-13-936435-8.
7. Shachak, A.; Kuziemsky, C.; Petersen, C. Beyond TAM and UTAUT: Future directions for HIT implementation research. *J. Biomed. Informatics* **2019**, *100*, 103315. https://doi.org/10.1016/j.jbi.2019.103315.
8. Gansser, O.A.; Reich, C.S. A new acceptance model for artificial intelligence with extensions to UTAUT2: An empirical study in three segments of application. *Technol. Soc.* **2021**, *65*, 101535. https://doi.org/10.1016/j.techsoc.2021.101535.
9. Holden, R.J.; Karsh, B.-T. The Technology Acceptance Model: Its past and its future in health care. *J. Biomed. Informatics* **2010**, *43*, 159–172. https://doi.org/10.1016/j.jbi.2009.07.002.
10. Ammenwerth, E. Technology Acceptance Models in Health Informatics: TAM and UTAUT. *Stud. Health Technol. Inform.* **2019**, *263*, 64–71. https://doi.org/10.3233/SHTI190111.
11. Turner, M.; Kitchenham, B.; Brereton, P.; Charters, S.; Budgen, D. Does the technology acceptance model predict actual use? A systematic literature review. *Inf. Softw. Technol.* **2010**, *52*, 463–479. https://doi.org/10.1016/j.infsof.2009.11.005.
12. Davis, F.D. Perceived Usefulness, Perceived Ease of Use, and User Acceptance of Information Technology. *MIS Q.* **1989**, *13*, 319–340. https://doi.org/10.2307/249008.
13. Venkatesh, V.; Davis, F.D. A Theoretical Extension of the Technology Acceptance Model: Four Longitudinal Field Studies. *Manag. Sci.* **2000**, *46*, 186–204. https://doi.org/10.1287/mnsc.46.2.186.11926.
14. Rondan-Cataluña, F.J.; Arenas-Gaitán, J.; Ramírez-Correa, P.E. A comparison of the different versions of popular technology acceptance models. *Kybernetes* **2015**, *44*, 788–805. https://doi.org/10.1108/k-09-2014-0184.
15. Yousafzai, S.Y.; Foxall, G.R.; Pallister, J.G. Technology acceptance: A meta-analysis of the TAM: Part 1. *J. Model. Manag.* **2007**, *2*, 251–280. https://doi.org/10.1108/17465660710834453.
16. Yousafzai, S.Y.; Foxall, G.R.; Pallister, J.G. Technology acceptance: A meta-analysis of the TAM: Part 2. *J. Model. Manag.* **2007**, *2*, 281–304. https://doi.org/10.1108/17465660710834462.
17. Andreoni, J. Giving with Impure Altruism: Applications to Charity and Ricardian Equivalence. *J. Polit. Econ.* **1989**, *97*, 1447–1458. https://doi.org/10.1086/261662.
18. Shakeri, A.; Kugathasan, H. Defining Donation. *Hektoen Int. J. Med. Humanit.* **2020**, *12*. Available online: https://hekint.org/2020/01/30/defining-donation/ (accessed on 5 September 2022).
19. Van de Ven, J. *The Economics of the Gift*; University of Amsterdam; Tinbergen Institute: Amsterdam, The Netherlands, 2000.
20. Saito, K. Impure altruism and impure selfishness. *J. Econ. Theory* **2015**, *158*, 336–370. https://doi.org/10.1016/j.jet.2015.05.003.
21. Saravanos, A.; Zheng, D.; Zervoudakis, S. Measuring Consumer Perceived Warm-Glow for Technology Adoption Modeling. Available online: https://arxiv.org/pdf/2203.09023.pdf (accessed on 19 March 2022).
22. Kuijpers, M.H. Perceptive Categories and the Standard of the Time. In *An Archaeology of Skill*; Routledge: London, UK, 2017; pp. 64–81, ISBN 1-315-19602-6.
23. Warshaw, P.R.; Davis, F.D. Disentangling behavioral intention and behavioral expectation. *J. Exp. Soc. Psychol.* **1985**, *21*, 213–228. https://doi.org/10.1016/0022-1031(85)90017-4.

24. Griskevicius, V.; Tybur, J.M.; Van den Bergh, B. Going green to be seen: Status, reputation, and conspicuous conservation. *J. Personal. Soc. Psychol.* **2010**, *98*, 392–404. https://doi.org/10.1037/a0017346.
25. Van den Bergh, B.; Griskevicius, V.; Tybur, J. Consumer Choices: Going Green to Be Seen. *RSM Discov. Manag. Knowl.* **2010**, *4*, 10–11.
26. Dastrup, S.R.; Zivin, J.G.; Costa, D.L.; Kahn, M.E. Understanding the Solar Home price premium: Electricity generation and "Green" social status. *Eur. Econ. Rev.* **2012**, *56*, 961–973. https://doi.org/10.1016/j.euroecorev.2012.02.006.
27. Hartmann, P.; Apaolaza-Ibáñez, V. Consumer attitude and purchase intention toward green energy brands: The roles of psychological benefits and environmental concern. *J. Bus. Res.* **2012**, *65*, 1254–1263. https://doi.org/10.1016/j.jbusres.2011.11.001.
28. Ma, C.; Burton, M. Warm glow from green power: Evidence from Australian electricity consumers. *J. Environ. Econ. Manag.* **2016**, *78*, 106–120. https://doi.org/10.1016/j.jeem.2016.03.003.
29. Sun, P.-C.; Wang, H.-M.; Huang, H.-L.; Ho, C.-W. Consumer attitude and purchase intention toward rooftop photovoltaic installation: The roles of personal trait, psychological benefit, and government incentives. *Energy Environ.* **2020**, *31*, 21–39. https://doi.org/10.1177/0958305x17754278.
30. Azalia, H.; Israni, H.; Indrasari, N.; Simamora, B.H. How Environmental Concern, Warm Glow, and Financial Impact Decision of Adopting Solar PV. *Int. J. Organ. Bus. Excell.* **2021**, *4*, 29–40.
31. Bhutto, M.Y.; Liu, X.; Soomro, Y.A.; Ertz, M.; Baeshen, Y. Adoption of Energy-Efficient Home Appliances: Extending the Theory of Planned Behavior. *Sustainability* **2021**, *13*, 250. https://doi.org/10.3390/su13010250.
32. Karjalainen, S.; Ahvenniemi, H. Pleasure is the profit—The adoption of solar PV systems by households in Finland. *Renew. Energy* **2019**, *133*, 44–52. https://doi.org/10.1016/j.renene.2018.10.011.
33. Tiger, L. *The Pursuit of Pleasure*; Routledge: London, UK, 2017; ISBN 1-315-13441-1.
34. Miltgen, C.L.; Popovič, A.; Oliveira, T. Determinants of end-user acceptance of biometrics: Integrating the "Big 3" of technology acceptance with privacy context. *Decis. Support Syst.* **2013**, *56*, 103–114. https://doi.org/10.1016/j.dss.2013.05.010.
35. Moore, G.C.; Benbasat, I. Development of an Instrument to Measure the Perceptions of Adopting an Information Technology Innovation. *Inf. Syst. Res.* **1991**, *2*, 192–222. https://doi.org/10.1287/isre.2.3.192.
36. Venkatesh, V. Determinants of Perceived Ease of Use: Integrating Control, Intrinsic Motivation, and Emotion into the Technology Acceptance Model. *Inf. Syst. Res.* **2000**, *11*, 342–365. https://doi.org/10.1287/isre.11.4.342.11872.
37. Kuruvatti, J.; Prasad, V.; Williams, R.; Harrison, M.A.; Jones, R.P.O. Motivations for donating blood and reasons why people lapse or never donate in Leeds, England: A 2001 questionnaire-based survey. *Vox Sang.* **2011**, *101*, 333–338. https://doi.org/10.1111/j.1423-0410.2011.01488.x.
38. Webster, J.; Martocchio, J.J. Microcomputer Playfulness: Development of a Measure with Workplace Implications. *MIS Q.* **1992**, *16*, 201–226. https://doi.org/10.2307/249576.
39. Abbey, J.D.; Meloy, M.G. Attention by design: Using attention checks to detect inattentive respondents and improve data quality. *J. Oper. Manag.* **2017**, *53-56*, 63–70. https://doi.org/10.1016/j.jom.2017.06.001.
40. Saravanos, A.; Zervoudakis, S.; Zheng, D.; Stott, N.; Hawryluk, B.; Delfino, D. The Hidden Cost of Using Amazon Mechanical Turk for Research. In Proceedings of the HCI International 2021–Late Breaking Papers: Design and User Experience, Washington DC, USA, 24-29 July 2021; Stephanidis, C., Soares, M.M., Rosenzweig, E., Marcus, A., Yamamoto, S., Mori, H., Rau, P.-L.P., Meiselwitz, G., Fang, X., Moallem, A., Eds.; Springer International Publishing: Cham, Switzerland, 2021; pp. 147–164.
41. Kock, N.; Hadaya, P. Minimum sample size estimation in PLS-SEM: The inverse square root and gamma-exponential methods. *Inf. Syst. J.* **2018**, *28*, 227–261. https://doi.org/10.1111/isj.12131.
42. Barclay, D.W.; Higgins, C.A.; Thompson, R. The Partial Least Squares Approach to Causal Modeling: Personal Computer Adoption and Use as an Illustration. *Technol. Stud.* **1995**, *2*, 284–324.
43. Chin, W.W.; Newsted, P.R. Structural Equation Modeling Analysis with Small Samples Using Partial Least Squares. In *Statistical Strategies for Small Sample Research*; Hoyle, R.H., Ed.; Sage Publications: Thousand Oaks, CA, USA, 1999.
44. Ringle, C.M.; Wende, S.; Becker, J.-M. *SmartPLS 3*; SmartPLS: Bönningstedt, Germany, 2015.
45. Islam, A.N.; Azad, N. Satisfaction and continuance with a learning management system: Comparing Perceptions of Educators and Students. *Int. J. Inf. Learn. Technol.* **2015**, *32*, 109–123. https://doi.org/10.1108/ijilt-09-2014-0020.
46. Chin, W.W. The Partial Least Squares Approach to Structural Equation Modeling. In *Modern Methods for Business Research*; Marcoulides, G.A., Ed.; Lawrence Erlbaum Associates Publishers: Mahwah, NJ, USA, 1998; Volume 295, pp. 295–336.
47. Fornell, C.; Larcker, D.F. Evaluating Structural Equation Models with Unobservable Variables and Measurement Error. *J. Mark. Res.* **1981**, *18*, 39–50. https://doi.org/10.2307/3151312.
48. Chin, W.W. Commentary: Issues and Opinion on Structural Equation Modeling. *MIS Q.* **1998**, *22*, vii–xvi.
49. Hair, J.F.; Ringle, C.M.; Sarstedt, M. PLS-SEM: Indeed a Silver Bullet. *J. Mark. Theory Pract.* **2011**, *19*, 139–152. https://doi.org/10.2753/mtp1069-6679190202.
50. Rigdon, E.E.; Sarstedt, M.; Ringle, C.M. On Comparing Results from CB-SEM and PLS-SEM: Five Perspectives and Five Recommendations. *ZFP–J. Res. Manag.* **2017**, *39*, 4–16. https://doi.org/10.15358/0344-1369-2017-3-4.
51. Jannoo, Z.; Yap, B.W.; Auchoybur, N.; Lazim, M.A. The Effect of Nonnormality on CB-SEM and PLS-SEM Path Estimates. *Int. J. Math. Comput. Phys. Quantum Eng.* **2014**, *8*, 285–291.
52. Haenlein, M.; Kaplan, A.M. A Beginner's Guide to Partial Least Squares Analysis. *Underst. Stat.* **2004**, *3*, 283–297. https://doi.org/10.1207/s15328031us0304_4.

17 of 17

53. Rigdon, E.E.; Ringle, C.M.; Sarstedt, M. Structural Modeling of Heterogeneous Data with Partial Least Squares. In *Review of Marketing Research*; Malhotra, N.K., Ed.; Review of Marketing Research; Emerald Group Publishing Limited: Bingley, UK, 2010; Volume 7, pp. 255–296, ISBN 978-0-85724-475-8.
54. Astrachan, C.B.; Patel, V.K.; Wanzenried, G. A comparative study of CB-SEM and PLS-SEM for theory development in family firm research. *J. Fam. Bus. Strat.* **2014**, *5*, 116–128. https://doi.org/10.1016/j.jfbs.2013.12.002.
55. Hair, J.F.; Risher, J.J.; Sarstedt, M.; Ringle, C.M. When to use and how to report the results of PLS-SEM. *Eur. Bus. Rev.* **2019**, *31*, 2–24. https://doi.org/10.1108/ebr-11-2018-0203.
56. Kock, N.; Lynn, G.S. Stevens Institute of Technology Lateral Collinearity and Misleading Results in Variance-Based SEM: An Illustration and Recommendations. *J. Assoc. Inf. Syst.* **2012**, *13*, 546–580. https://doi.org/10.17705/1jais.00302.
57. Hagedoorn, J.; Wang, N. Is there complementarity or substitutability between internal and external R&D strategies? *Res. Policy* **2012**, *41*, 1072–1083. https://doi.org/10.1016/j.respol.2012.02.012.
58. Henseler, J.; Ringle, C.M.; Sinkovics, R.R. *The Use of Partial Least Squares Path Modeling in International Marketing*; Sinkovics, R.R., Ghauri, P.N., Eds.; Emerald Group Publishing Limited: Bingley, UK, 2009; p. 319, ISBN 978-1-84855-468-9.
59. Hu, L.T.; Bentler, P.M. Cutoff criteria for fit indexes in covariance structure analysis: Conventional criteria versus new alternatives. *Struct. Equ. Model. Multidiscip. J.* **1999**, *6*, 1–55. https://doi.org/10.1080/10705519909540118.
60. Henseler, J.; Ringle, C.M.; Sarstedt, M. A new criterion for assessing discriminant validity in variance-based structural equation modeling. *J. Acad. Mark. Sci.* **2015**, *43*, 115–135. https://doi.org/10.1007/s11747-014-0403-8.
61. Henseler, J.; Hubona, G.; Ray, P.A. Using PLS path modeling in new technology research: Updated guidelines. *Ind. Manag. Data Syst.* **2016**, *116*, 2–20. https://doi.org/10.1108/imds-09-2015-0382.
62. Akaike, H. A new look at the statistical model identification. *IEEE Trans. Autom. Control* **1974**, *19*, 716–723. https://doi.org/10.1109/tac.1974.1100705.
63. Venkatesh, V.; Morris, M.G.; Davis, G.B.; Davis, F.D. User Acceptance of Information Technology: Toward a Unified View. *MIS Q.* **2003**, *27*, 425–478. https://doi.org/10.2307/30036540.
64. King, W.R.; He, J. A meta-analysis of the technology acceptance model. *Inf. Manag.* **2006**, *43*, 740–755. https://doi.org/10.1016/j.im.2006.05.003.
65. Venkatesh, V.; Thong, J.Y.L.; Xu, X. Consumer Acceptance and Use of Information Technology: Extending the Unified Theory of Acceptance and Use of Technology. *MIS Q.* **2012**, *36*, 157–178. https://doi.org/10.2307/41410412.
66. Al-Gahtani, S.S.; Hubona, G.S.; Wang, J. Information technology (IT) in Saudi Arabia: Culture and the acceptance and use of IT. *Inf. Manag.* **2007**, *44*, 681–691. https://doi.org/10.1016/j.im.2007.09.002.
67. Bandyopadhyay, K.; Fraccastoro, K.A. The Effect of Culture on User Acceptance of Information Technology. *Commun. Assoc. Inf. Syst.* **2007**, *19*, 23. https://doi.org/10.17705/1cais.01923.
68. Faqih, K.M.S.; Jaradat, M.-I.R.M. Assessing the moderating effect of gender differences and individualism-collectivism at individual-level on the adoption of mobile commerce technology: TAM3 perspective. *J. Retail. Consum. Serv.* **2015**, *22*, 37–52. https://doi.org/10.1016/j.jretconser.2014.09.006.
69. Im, I.; Hong, S.; Kang, M.S. An international comparison of technology adoption: Testing the UTAUT model. *Inf. Manag.* **2011**, *48*, 1–8. https://doi.org/10.1016/j.im.2010.09.001.
70. Srite, M.; Karahanna, E. The Role of Espoused National Cultural Values in Technology Acceptance. *MIS Q.* **2006**, *30*, 679–704. https://doi.org/10.2307/25148745.
71. Yuen, Y.Y.; Yeow, P.; Lim, N. Internet banking acceptance in the United States and Malaysia: A cross-cultural examination. *Mark. Intell. Plan.* **2015**, *33*, 292–308. https://doi.org/10.1108/mip-08-2013-0126.
72. Iweala, S.; Spiller, A.; Meyerding, S. Buy good, feel good? The influence of the warm glow of giving on the evaluation of food items with ethical claims in the U.K. and Germany. *J. Clean. Prod.* **2019**, *215*, 315–328. https://doi.org/10.1016/j.jclepro.2018.12.266.